\documentclass[aps,prl,twocolumn,showpacs,amsmath,amsfonts]{revtex4}
\usepackage{graphicx}
\begin{document}

\title{Electric field induced nucleation: An alternative pathway to metallic hydrogen}
\date{\today}

\author{M. Nardone} \affiliation{Department of Physics and Astronomy, University of Toledo, Toledo, OH 43606, USA}
\author{V. G. Karpov} \affiliation{Department of Physics and Astronomy, University of Toledo, Toledo, OH 43606, USA}

\begin{abstract}
Electric field induced nucleation is introduced as a possible mechanism to realize a metallic phase of hydrogen. Analytical expressions are derived for the nucleation probabilities of both thermal and quantum nucleation in terms of material parameters, temperature, and the applied field.  Our results show that the insulator-metal transition can be driven by an electric field within a reasonable temperature range and at much lower pressures than the current paradigm of $P\gtrsim 400$ GPa.  Both static and oscillating fields are considered and practical implementations are discussed.
\end{abstract}

\maketitle

Metallic hydrogen (MH) was predicted by Wigner and Huntington \cite{wigner1935} in 1935 and it continues to be on the forefront of scientific inquiry; new predictions include high-$T$ superconductivity \cite{ashcroft1968, cudazzo2008} and novel ordered quantum fluid behavior \cite{ashcroft2000, babaev2004}.  A significant body of experimental work \cite{experimental, eremets2009, deemyad2008, boriskov2010}  and advanced numerical simulations \cite{numerical} have elucidated the properties of hydrogen at high pressure and have provided a great deal of insight to its rich phase diagram \cite{silvera2010}. From a practical standpoint, if metastable MH \cite{brovman1972} were to be obtained, it could be a revolutionary clean energy source \cite{silvera2010a}.

Attempts to produce MH have focused on high pressure techniques.  Solid MH has not yet been observed under static pressures of up to 342 GPa \cite{narayana1998, loubeyre2002}.  Dynamic compression beyond 200 GPa has also been employed \cite{hicks2009, boriskov2010}.  The only direct evidence so far was the brief observation of a highly conductive liquid phase under a shockwave pressure of 140 GPa and temperatures around 3000 K \cite{weir1996}.  Recent numerical studies have calculated a solid \cite{stadele2000, johnson2000} metal-insulator transition pressure $\gtrsim$ 400 GPa and a first order liquid-liquid \cite{lorenzen2010} metal-insulator transition pressure $\sim$ 200 GPa near 1000 K.

In this work, we introduce electric field induced nucleation (FIN) as a possible alternative means by which to realize MH;  both thermal and quantum FIN are considered.  The underlying thermodynamic theory is general in nature and not concerned with the microscopic aspects of the phase transition, whether it occurs by dissociation of the molecular form to a monatomic metal, as originally considered \cite{wigner1935}, or from a proton-paired insulator to a molecular metallic state \cite{johnson2000}.  Beyond MH, FIN could open new venues for phase transformations to otherwise unobtainable materials.

\begin{figure}[htb]
\includegraphics[width=0.45\textwidth]{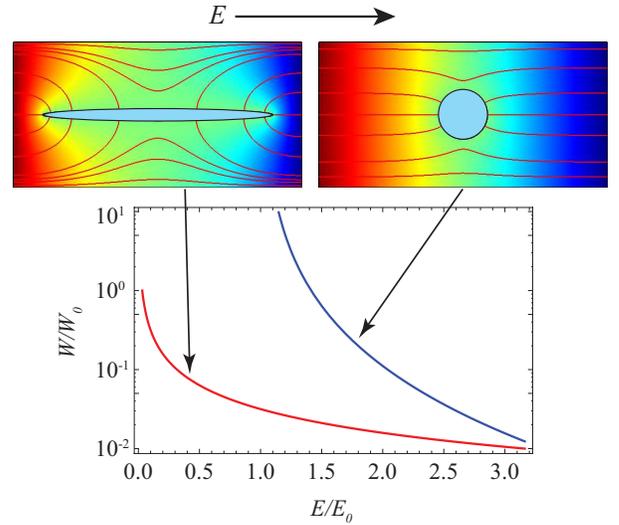}
\caption{Electric potential surface plot and field streamlines show the greater electrostatic energy reducing effect of elongated metallic nuclei versus spheres.  The corresponding decrease in the the nucleation barrier $W/W_0$ with respect to the applied field $E/E_0$ is shown in the lower plot.\label{Fig:fields}}
\end{figure}

FIN is a recently developed concept of metal phase nucleation in an insulating host under a strong static \cite{karpov2008} or oscillating \cite{karpov2010} field.  FIN produces needle-shaped conductive embryos with vertical axes aligned with the field (see Fig. \ref{Fig:fields}).  It has been identified as a mechanism for crystallization in chalcogenide glasses \cite{karpov2008}.  Also, it is supported by observations of laser induced nucleation of particles aligned with the beam polarization \cite{NPIN}.

Given the high electrical conductivity of a metallic nucleus, the field $E$ causes it to act as a strong dipole, $p\propto E$.  The associated negative contribution, $-pE$, to the free energy suppresses the nucleation barrier.  The premise of FIN is that anisotropic particles can result in much greater barrier suppression than spheres, as shown in Fig. \ref{Fig:fields}.  Specifically, a metal particle of volume $\Omega$ in a uniform field $E$ reduces the free energy according to \cite{landau1984},
\begin{equation}\label{eq:elec}
F_E=-\frac{\varepsilon E^2}{8\pi n}\Omega,
\end{equation}
where $\varepsilon$ is the permittivity of the insulating host and $n$ is the depolarizing factor, which depends on the particle geometry (see Fig. \ref{Fig:fields}).  For example, a sphere provides $n=1/3$, while a prolate spheroid of radius $R$ and height $H$ yields $n=(R/H)^2[\ln(2H/R)-1]$. An approximate relation $n\sim (R/H)^2$ holds for arbitrary needle-shaped particles. Indeed, the condition that the dipole field balances the external field in a metal determines $p\sim EH^3$, leading to $F_E\sim -pE \sim E^2H^3\sim -\Omega E^2(H/R)^2$ with $\Omega\sim R^2H$.

According to FIN, the free energy of a MH nucleus can be expressed as,
\begin{equation}\label{eq:free}
F=A\sigma +\Omega|\mu|+F_E,
\end{equation}
where $A$ is its surface area, $\sigma$ is the surface tension, and $\mu$ is the chemical potential difference between the two phases.  The positive chemical potential term in Eq. (\ref{eq:free}) corresponds to a system that is stable in zero field.

\begin{figure}[htb]
\includegraphics[width=0.48\textwidth]{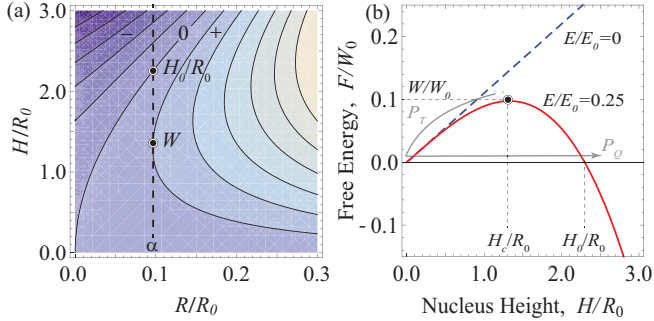}
\caption{(a) Contours of free energy from Eq. (\ref{eq:freecyl}); positive and negative regions separated by the zero contour.  Nucleation of elongated particles of critical height $H_c/R_0$ is governed by the barrier, $W$. (b) Cross-section of the free energy from (a) along the line $\alpha$ shown with (solid) and without (dashed) the electric field $E/E_0$.  Thermal and quantum nucleation pathways are labeled $P_T$ and $P_Q$. \label{Fig:contour}}
\end{figure}

The exact shape of the elongated nuclei is not known, but modeling with either spheroidal or cylindrical particles leads to differences only in numerical coefficients \cite{karpov2008}.  We opt for the mathematically more concise form of a cylindrical nucleus with the free energy of Eq. (\ref{eq:free}) given by,
\begin{equation}\label{eq:freecyl}
F_{cyl}=\frac{W_0}{2}\left(\frac{3RH}{R_0^2}+\frac{3R^2H}{R_0^3}-\frac{E^2}{E_0^2}\frac{H^3}{R_0^3}\right).
\end{equation}
Here, we have approximated the logarithmic term in Eq. (\ref{eq:elec}) with unity, and introduced the nucleation barrier, $W_0=16\pi\sigma^3/(3\mu^2)$, and radius, $R_0=2\sigma/|\mu|$, for spherical particles of the classical nucleation theory; the characteristic field is $E_0= 2\sqrt{W_0/\varepsilon R_0^3}$. In Fig. \ref{Fig:contour}, a contour plot of the free energy illustrates how the system can lower its free energy by forming elongated particles.

The free energy of Eq. (\ref{eq:freecyl}) seems to suggest that nuclei with $R\rightarrow 0$ are energetically favorable.  Realistically, $R$ must be greater than some minimum value determined by extraneous requirements, such as sufficient conductivity to support a large dipole energy or mechanical integrity. Based on data for other types of systems, it was estimated \cite{karpov2008} that a reasonable minimum radius is $R_{min}=\alpha R_0$, where $\alpha\sim 0.1$. Lacking other data we maintain $\alpha\sim 0.1$, which leads to $R_{min}$ on the order of interatomic distance.  The free energy in the region $R/R_0<\alpha$ is substantially larger than described by Eq. (\ref{eq:freecyl}), since the energy reducing effect of the electric field cannot be manifested by such small particles.  As a simplifying approximation we consider nucleation along the path  $R/R_0=\alpha$ (see Fig. \ref{Fig:contour}); alternative paths that start from the point (0,0) introduce only insignificant numerical factors.

From Eq. (\ref{eq:freecyl}), the nucleation barrier is,
\begin{equation}\label{eq:barrier}
W=(E_c/E)W_0, \quad E_c\equiv \alpha ^{3/2}E_0.
\end{equation}
Thermally activated FIN, with the probability $P_T\propto \exp(-W/kT)$, becomes important when $E>E_c\equiv\alpha^{3/2}E_0$.  The requirement of an elongated nucleus, $H/R\gg 1$, places an upper limit on the field, $E<E_c/\alpha^2$.

For MH, using \cite{brovman1972a} $\mu\sim 0.1$ Ry/atom $\sim 10^{11}$ J/m$^3$ and $\sigma \sim 1$  Ry/atom (for structurally different phases) yields $R_0\sim 10$ $\textrm{\AA}$ and $W_0\sim \mu R_0^3\sim 10^3$ eV.  Then with $\alpha=0.1$, and \cite{boriskov2010} $\varepsilon\sim 1$, we obtain $E_c\sim 10^7$ V/cm.


Following the approaches in Refs. \onlinecite{lifshitz1972, widom1984}, quantum nucleation has the probability exponent,
\begin{equation}\label{eq:barrierfactor}
S=\frac{2}{\hbar}\int\limits_{C}\!\sqrt{2M(R,H)F_{cyl}(R,H)}\,\mathrm{d}s,
\end{equation}
where $M(R,H)$ is the local mass.  The integral is taken over the tunneling path $C$ with element $\mathrm{d}s$ through the two-dimensional free energy barrier.

To determine $M$, we consider a host material of density $\rho$ in which a cylindrical nucleus of density $\rho+\Delta\rho$ grows by the inward flux of particles with radial speed,
\begin{equation}\label{eq:velocity}
v_r=-\frac{\Delta\rho}{\rho}\frac{R}{r}\dot{R}\quad \mathrm{for}\quad H\gg r>R,
\end{equation}
where $\dot{R}$ is the radial growth velocity of the nucleus.
The associated hydrodynamic kinetic energy takes the form,
\begin{equation}\label{eq:kinetic}
E_k\approx (\rho/2)\int_R^{H}\!v_r^2(2\pi H)\,r\mathrm{d}r = M\dot{R}^2/2,
\end{equation}
from which,
\begin{equation}\label{eq:mass}
M(R,H)=2\pi R^2H\frac{\Delta\rho^2}{\rho}\ln\left(\frac{H}{R}\right).
\end{equation}
Here, we have taken into account that the assumption of an axially symmetric velocity field in Eq. (\ref{eq:velocity}) is valid when $r\ll H$.  The region $r\gg H$ is closer to spherical symmetry (considered in Refs. \cite{lifshitz1972, widom1984}) where the integral for kinetic energy is dominated by its lower limit and is relatively insignificant.

Inserting Eqs. (\ref{eq:freecyl}) and (\ref{eq:mass}) into Eq. (\ref{eq:barrierfactor}) and using dimensionless variables $x=R/R_0$ and $y=H/R_0$, results in the barrier factor,
\begin{equation}\label{eq:barrier2}
S=S_0\int\limits_{C}\!\left[x^3y^2\ln{\left(\frac{y}{x}\right)}\left(1+x-\frac{E^2y^2}{3xE_0^2}\right)\right]^{1/2}\,\mathrm{d}s,
\end{equation}
where $S_0=(6\pi \Delta \rho ^2R_0^5W_0/\rho \hbar^2)^{1/2}$ and the integral is dimensionless. Optimizing the latter leads to an equation of motion for the optimal path that is analytically intractable.  We limit our approximation to the sufficiently simple but physically reasonable path along the line $x=\alpha\ll 1$, as shown in Fig. \ref{Fig:contour}. The path ends at the point $H_0/R_0 = \sqrt{3}E_c/(\alpha E)$, which follows from $F_{cyl}(R_{min},H_0)=0$.  Evaluation of Eq. (\ref{eq:barrier2}) then yields the quantum nucleation probability $P_Q\propto\exp{(-S)}$,
\begin{equation}\label{eq:probquantum}
P_Q\propto\exp{\left[-S_0\left(\frac{E_c}{\gamma E}\right)^2\right]},
\end{equation}
where $\gamma=\left[\alpha/\ln{\left(\sqrt{3}E_c/\alpha ^2E\right)}\right]^{1/4}$.  The field-dependent expression in parentheses describes the FIN effect; setting that expression to unity reduces Eq. (\ref{eq:probquantum}) to that derived in Ref. \onlinecite{lifshitz1972} (to within insignificant numerical factors).
Several other pathways through the barrier differing from the line $x=\alpha$ were found to result in larger values for the barrier factor, making our approximation along the line $x=\alpha$ a reasonable estimate of the variational minimum.

\begin{figure}[htb]
\includegraphics[width=0.48\textwidth]{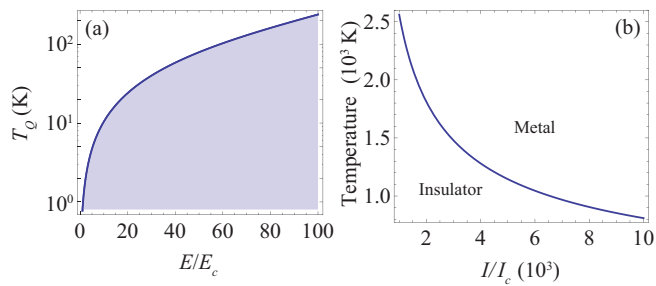}
\caption{(a) Temperature regime of quantum nucleation ($T<T_Q$) as a function of applied field $E/E_c$. (b) An approximate $T-E$ phase diagram of field induced nucleation (with the field given by the relative laser intesity $E/E_c=\sqrt{I/I_c}$) . \label{Fig:temps}}
\end{figure}

Neglecting any difference in the probability pre-exponentials, the temperature of interplay between activated and quantum nucleation is given by
\begin{equation}\label{eq:temp}
T_Q=\frac{W_0}{kS_0}\frac{\gamma^2 E}{E_c}.
\end{equation}
If, for example, we consider the solid molecular to solid metallic transition near $T=0$, then \cite{wigner1935} $\Delta\rho/\rho\sim 10$ and $\Delta\rho\sim 1$ g/cm$^3$.  With the previous values of $W_0\sim 10^3$ eV and $R_0\sim 10$ $\textrm{\AA}$, the temperature regime of quantum nucleation is shown in Fig. \ref{Fig:temps}(a).

Our treatment thus far equally applies to the case of oscillating laser fields with the equivalent static field given by $E=\sqrt{4\pi I/c}$, where $I$ is the beam intensity and $c$ the speed of light.  As detailed in Ref. \onlinecite{karpov2010}, the equivalent static field can be applied under the following conditions: 1) the radius of the metal particle less than the field decay length, $R<\delta=c\sqrt{1/2\pi\sigma\omega}$, where $\omega$ is the laser frequency and $\sigma$ is the MH conductivity; and 2) the MH particle must be sufficiently polarizable to ensure a strong field effect.  The latter condition can be expressed \cite{karpov2010},
\begin{equation}\label{eq:freq}
\frac{\omega_p}{\omega}\frac{E}{E_0}\gg\sqrt{\frac{1}{\alpha}},
\end{equation}
where $\omega_p$ is the MH plasma frequency.  With adherence to these conditions, all of the above results hold with the substitution $E/E_c\rightarrow \sqrt{I/I_c}$ where,
\begin{equation}\label{eq:intensity}
I_c=\frac{cE_c^2}{4\pi}=\frac{cW_0\alpha^3}{\pi \varepsilon R_0^3}\sim 10^{12}\; \mathrm{W/cm}^2.
\end{equation}

Optical absorption can facilitate a temperature increase $\delta T$ above ambient $T_0$ so that $T=T_0+\delta T$. In terms of material parameters, the activated and quantum nucleation rates can be estimated respectively as,
\begin{equation}\label{eq:ratethermal}
P_T=N_0\omega_0\exp{\left[-\frac{W_0}{k\left(T_0+\delta T\right)}\sqrt{\frac{I_c}{I}}\right]},
\end{equation}
and
\begin{equation}\label{eq:ratequantum}
P_Q=N_0\omega_0\exp\left(-S_0\frac{I_c}{\gamma^2 I}\right).
\end{equation}
In the pre-exponential factor $N_0\omega_0\sim 10^{35}$ cm$^{-3}$s$^{-1}$, $N_0\sim R_0^{-3}$ is the number density of fluctuating centers and $\omega _0\sim W_0/(S_0\hbar )$ is the characteristic vibrational frequency of such a center.

For $T\gg T_Q$, by setting $P_T=1$ in Eq. (\ref{eq:ratethermal}) we obtain a reasonable transition temperature as a function of the applied field (or intensity).  The result is an approximate $T-E$ phase diagram, as shown in Fig. \ref{Fig:temps}(b), which indicates that thermal FIN can be practical.

Given the very rough numerical values above, we find that the probability of quantum nucleation is negligible, even at the maximum allowable intensity of $I\sim 10^4 I_c$ (or static field $E=E_c/\alpha^2\sim 10^2 E_c$).  However, our analysis is based on quantum nucleation at $T=0$ and ambient pressure.  At higher temperatures, nucleation may be more efficient through an optimum combination of activation and tunneling, especially considering the broad temperature regime of quantum nucleation [see Fig. \ref{Fig:temps}(a)]. In addition, since our numerical estimates are based on solid molecular to solid metallic transitions at atmospheric pressure, phase transitions at higher pressures and temperatures, such as solid-liquid or liquid-liquid, would occur with lower barriers $W_0$ and density ratios $\Delta\rho/\rho$, leading to shorter nucleation times.


One conceivable implementation of FIN is related to recent diamond anvil cell experiments wherein cryogenically loaded hydrogen was heated with either continuous wave \cite{eremets2009} or pulsed \cite{deemyad2008} lasers.  In those experiments, which were designed to investigate the melting line of hydrogen up to nearly 150 GPa, a laser of wavelength $\lambda\sim 1$ $\mu$m and beam intensity $I\sim 10^6-10^9$ W/cm$^2$ was focused to a spot size of 10-30 $\mu$m on a Pt absorber located within the hydrogen sample.  These melting experiments could be slightly modified to assess the FIN theory by increasing the laser intensity (i.e. increase power and/or reduce area) to the range of $I\gtrsim I_c$ while maintaining the frequency required by Eq. (\ref{eq:freq}); given $\omega_p\sim 4\times10^{16}$ s$^{-1}$ for hydrogen. Metal nucleation could be detected by changes in, for example, conductivity, opacity and/or scattering within the portion of the sample subtended by the beam.  Since the electric field is the primary phase change driver, the experiments could be conducted at much lower pressures and over a broad range of temperatures.  Simultaneously, one could investigate the correlation between nucleus orientation and beam polarization.  Similar experiments could be conducted with hydrogen rich alloys such as CH$_4$ (or other paraffins) and SiH$_4$(H$_2$)$_2$ \cite{ashcroft2004, stobel2009, yao2010, eremets2008}, or LiH$_n$ ($n>1$) \cite{zurek2009}.


In conclusion, our results show that FIN could present a new pathway to metallic hydrogen that is within reach of existing experimental techniques.  Further investigation is imperative since there is theoretical evidence \cite{brovman1972} that the metallic phase would exhibit long-term metastability upon removal of the field at ambient pressure, thereby providing a novel source of clean energy.  In addition, FIN may be useful in realizing other materials that have been thus far unobtainable.  

\end{document}